\documentclass{aa}
\usepackage{graphics}
\begin{document}
 
\thesaurus{02.(02.09.1;02.08.1)08.(08.09.3)}
 
\title{
Shear layer instability in a highly diffusive stably stratified atmosphere} 
\author{F. Ligni\`eres \inst{1,} \inst{2} \and F. Califano \inst{3}
\and A. Mangeney \inst{1}}
\institute{D\'epartement de Recherche Spatiale et Unit\'e de Recherche
associ\'ee au CNRS 264, Observatoire de Paris-Meudon, F-92195
Meudon Cedex, France \and
Astronomy Unit, Queen Mary \& Westfield College, Mile
End Road, London E14NS, UK \and
Istituto Nazionale Fisica della Materia, Sez. A, 
Universit\`a di
Pisa, Italy}
\offprints{F. Ligni\`eres}
\mail{F.Lignieres@qmw.ac.uk}
\date{Received  1 July 1999 / Accepted 5 August 1999}
\authorrunning
\titlerunning
\maketitle
\begin{abstract} 
The linear stability of a shear layer in a highly diffusive stably stratified atmosphere has been 
investigated. This study completes and extends previous works by Dudis (1974) and Jones (1977). 
We show that (i) an asymptotic regime is reached in the limit of 
large thermal diffusivity, 
(ii) 
there exists two different types of unstable modes.
The slowest growing modes correspond to predominantly horizontal motions and 
the fast 
growing ones to isotropic motions, 
(iii) the physical interpretation of the stability 
properties
in this asymptotic regime is
simple.
Applications to the dynamics of stellar radiative zone are discussed. 

\keywords{Instabilities -- hydrodynamics -- Stars:
interiors}
\end{abstract}

\section{Introduction}

The standard model describing stellar evolution
neglects the role of macroscopic motions in
the radiative layers of stars.
During the past twenty years, however, determinations of surface
chemical abundances and helioseismology data have shown that
some mechanical mixing must be invoked to account
for the observations.
Current mixing theories involve either motions induced by the rotation
or gravity waves generated 
at the boundary with the convective zones.

Hydrodynamical instabilities due to differential rotation
play an essential role in this context. 
Turbulent motions arise through the non-linear
development of hydrodynamical instabilities and whenever present,
they ensure an efficient mixing of
chemical and angular momentum. The study of 
instabilities is therefore necessary
to determine the occurrence of turbulent mixing in stellar 
radiative zones.
Moreover, in the absence of a better understanding of turbulent motions, 
the properties of the instability are often used 
to infer the properties of the subsequent turbulent motions.

In a rotating fluid of uniform density, a hydrodynamical instability occurs if
the specific angular momentum decreases outwards the rotation axis or
if the rotational velocity varies along the rotation axis.
Instability can also be triggered by a velocity shear, a typical example
being the Kelvin-Helmholtz instability
which develops at the interface between two streams of different
velocities (see for example the reference books by
Chandrasekhar 1961 and Drazin 1981). In principle, these different types
of instability can also take
place inside stellar radiative zones.
But there, the stable stratification will
strongly oppose to them. Indeed, the time scale characterizing the action of the
restoring
buoyancy force
is generally much smaller than the dynamical time scale $\Omega^{-1}$
(the inverse of the Brunt-V\"{a}is\"{a}l\"{a}
frequency is of the order of $10^{3} \mbox{\rm s}$ whereas
$\Omega^{-1}$ is equal to
$10^{5.5}  \mbox{\rm s}$ for the sun and
is of the order of $10^{4} \mbox{\rm s}$
for
a fast rotating star with an equatorial velocity $v_{\rm eq} = 200 \mbox{\rm km
s}^{-1}$
and a radius $R=3R_{\odot}$).
 
This does not mean that instabilities are systematically
suppressed by stable stratification. But, as we shall see, this puts strong
constraints on
the type of motions
which can arise from these instabilities. 

A possibility to avoid the action of the buoyancy force is to consider displacements limited 
to the surfaces of constant entropy. 
Such purely two-dimensional motions are not affected by 
the buoyancy force and might be amplified 
if a differential rotation exists along these surfaces. 
However, even if the early 
development of the instability may not be 
affected by the
stabilizing buoyancy force, the non-linear 
stage should produce three-dimensional
motions
which in turn will be affected by the stable stratification.

Another possibility is to consider 
three-dimensional perturbations whose vertical length scale is 
small enough to be significantly 
affected by the thermal diffusion. Thermal diffusion has indeed 
a destabilizing effect in stably stratified atmospheres 
as was first recognized 
in a geophysical
context by Townsend (1958). When a fluid parcel is displaced vertically from its
hydrostatic
equilibrium position, 
thermal diffusion reduces the temperature departure between
the fluid parcel and its environment. For incompressible motions,
the density departure and so
the amplitude of the restoring buoyancy force are reduced by the same
factor. Thermal
diffusion
has therefore
a destabilizing effect because
it weakens the stabilizing effect of the stable
stratification.
 
In the present paper, we investigate in details this destabilizing effect 
in the case of shear instabilities.
Our work follows previous theoretical studies which we summarize now.
Let's first recall that, in the absence of diffusive and viscous effects,
there exists 
a criterion, derived in the context of the linear stability theory
and valid for any parallel flows ${\bf U} =U(z)
{\bf e}_x$
embedded in some vertically stably stratified Boussinesq atmosphere
($z$ is the vertical coordinate and
${\bf e}_x$ a horizontal unit vector).
This criterion established by Miles (1961) states
that the flow is stable to infinitesimal
perturbations if the Richardson number defined as,
\begin{equation} \label{eq:mi}
Ri = \left(\frac{N}{\frac{dU}{dz}} \right)^{2},
\end{equation}
\noindent
is everywhere larger than $1/4$.

Such a criterion does not exist 
in the presence of thermal diffusion.
Existing results consist either in
general criteria based on phenomenological
arguments or applications of the linear stability theory 
to a particular 
flow i.e. a particular velocity profile within a particular atmosphere.

The general criterion derived by Zahn (1974) states
that the shear layer is stable if,
\begin{equation} \label{eq:za}
Ri Pe > \frac{1}{4},
\end{equation}
\noindent
where $Pe= u l / \kappa $ is a P\'eclet number associated with 
an eddy of size $l$ and velocity $u$.
As expected, perturbations that would be 
stable in a perfect fluid because $Ri > 1/4$
can now be unstable provided $Pe < 1/(4 Ri)$, i.e. if 
the thermal 
diffusion time scale of the eddy is small enough compared to
its turnover time scale $l/u$.
More recently, Maeder (1995) obtained a similar expression by extending
to the diffusive case the energy considerations which allowed
Chandrasekhar's derivation of Miles' criterion
(Chandrasekhar 1961, see also Miles 1986).

In the context of the linear stability theory, thermal diffusion effects have
been studied independently by Dudis (1974) and Jones (1977). 
Both authors have chosen the hyperbolic-tangent velocity profile,
$U(z) = \Delta U \tanh (z/L)$, whose
instability properties 
are well-known in the unstratified case (Drazin 1981).
However they made different choices for the
temperature profile characterizing the stable stratification. 

Dudis considered a 
hyperbolic-tangent profile for the temperature $T(z) = T_0 +
\Delta T \tanh (z/L)$. He found in particular
that, for small enough P\'eclet numbers,
the stability depends only on the product of the Richardson
and P\'eclet number $Ri Pe$ and not on $Ri$ and $Pe$ separately,
this result being consistent with the criterion proposed by Zahn (1974) 
(here $Ri = (N  L)^2 / \Delta U^2$
and $Pe = \Delta U L/\kappa$).
However, it should be noticed that 
the thermal stratification chosen by Dudis is not
adapted to investigate the small P\'eclet number 
regime. Actually, in this regime,
the diffusion of the hyperbolic-tangent temperature profile is much faster 
than the growth of the instability, and this invalidates a starting assumption 
of this type of instability study, namely
that the basic state is a stationary solution of the governing equations 
or, at least, can be considered as 
such during the development of the instability.
The minimum growth time of the Kelvin-Helmholtz instability 
(i.e. the inverse of the maximum growth rate)
is known to be 
$\simeq 5 L / \Delta U $ whereas the thermal diffusion time scale 
of the temperature profile
is $t_{\kappa} = L^2 / \kappa$. Thus, if the P\'eclet number $Pe = 
\Delta U L/\kappa$
is smaller than unity, the basic temperature profile would have been diffused
much before the instability had time to grow. 

To avoid
this problem, one must consider 
a stationary solution of the heat equation and, 
for a Boussinesq fluid, this implies a temperature
profile varying
linearly with altitude.
Jones (1977) chose such a profile and then studied
the stability
in the range
$1/4 \leq Ri \leq 1$ as well as in the limit
of very large horizontal wavelengths of the disturbance, $k_x$,
and very large Richardson number. The stability criterion in 
this double 
limit is $k_x Pe Ri > 0.086$.
However,
the partial coverage of the parameter space ($k_x$, $Pe$, $Ri$) did not allow
to investigate the small P\'eclet number regime, and in particular,
the existence of the asymptotic state suggested by Dudis (1974).

Here we present a normal mode stability analysis of a hyperbolic-tangent
shear layer embedded in a linearly stably stratified Boussinesq atmosphere,
with the objective of completing the work of Dudis (1974) and Jones (1977).
First, we determine
the conditions of marginal stability
for a much larger range
of Richardson numbers, P\'eclet numbers and Reynolds numbers.
Then, we concentrate on
the small P\'eclet number regime and investigate for the first time 
the growth rates of the unstable modes 
as well as their spatial properties.
We also provide simple physical interpretation of the stability properties
by comparing the 
relative effects of shear, stable stratification and thermal diffusion.
Ligni\`eres (1999) recently derived 
an asymptotic form of the Boussinesq equations in the limit 
of small P\'eclet number and showed that, in this asymptotic regime, 
thermal diffusion and stable stratification combine in a single process.
We shall see that this property greatly simplifies 
the interpretation of our results.

The paper is organized as follows. The equations governing the 
evolution of perturbations are derived in Sect. 3.1 and the numerical method
to solve the corresponding eigenvalue problem is presented in Sect. 3.2.
Results are presented and interpreted in Sect. 4. In Sect. 5, they are
summarized and discussed in an astrophysical context.

\section{The mathematical model}

\subsection{Governing equations}

From now on, dimensional quantities are denoted with an overbar.
We consider a parallel flow ${\bar {\bf U}} = \overline{\Delta U} 
\tanh ({\bar z}) {\bf e}_{x}$ in an
unbounded 
vertically stratified atmosphere.
The ${\bar z}$ axis
refers to the direction of gravity, while the ${\bar x}$ axis corresponds
to the stream
direction. The equation of state is that of a Boussinesq fluid,
${\bar \varrho} = {\bar \varrho}_0 ( 1 - {\bar \beta} 
( {\bar T} - {\bar T}_0 ))$ and the thermal
stratification is given by the conductive
profile ${\bar T}_{\rm eq}(z) = {\bar T}_0 + 
\overline{\Delta T} {\bar z}/{\bar L}$.
Here ${\bar \beta}$
is the thermal expansion coefficient and,
${\bar \varrho}_0$ and ${\bar T}_0$ are references density and temperature.

Using,
${\bar L}$, $\overline{\Delta U}$ and $\overline{\Delta T}$ as units of
length, velocity and temperature, respectively,
the non-dimensional Boussinesq equations (Spiegel \& Veronis
1960) read:

\begin{equation} \label{eq:vel1}
\frac{\partial {\bf u}}{\partial t} + {\bf u}\cdot \nabla {\bf u} =
- \nabla p + Ri  \theta {\bf e}_z + \frac 1{Re} \nabla ^2{\bf u},
\end{equation}

\begin{equation} \label{eq:temp1}
\frac{\partial \theta}{\partial t} + {\bf u}\cdot \nabla \theta + w = 
\frac1{Pe}\nabla ^2 \theta,
\end{equation}

\begin{equation} \label{eq:div1}
\nabla \cdot {\bf u}=0,
\end{equation}
\noindent
where, ${\bf u} = u {\bf e}_{x} + v {\bf e}_{y} + w {\bf e}_{z}$
is the velocity vector, $p$
the pressure and $\theta(x,y,z) = T(x,y,z) - T_{\rm eq}$
the temperature deviation from the 
temperature profile $T_{\rm eq}(z)$. 
In the heat equation, the third term of the left
hand side
corresponds to the vertical advection of temperature against the mean
temperature gradient $d T_{\rm eq}(z)/dz$ which is
equal to unity in our
dimensionless units.

The system is governed by three non-dimensional numbers,
the Richardson number, $Ri$, the Reynolds number, $Re$, and the P\'eclet
number, $Pe$, respectively defined as
\[ Ri  = \left(\frac{{\bar N} {\bar L}}{\overline{\Delta U}}\right)^{2},
\;\;\;
Re = \frac{\overline{\Delta U} {\bar L}}{{\bar \nu}} ,\;\;\;
Pe = \frac{\overline{\Delta U} {\bar L}}{{\bar \kappa}},\]
\noindent
where, $N = \left({\bar \beta} {\bar g} \overline{\Delta T}/ {\bar L}
\right)^{1/2}$ is the
Brunt-V\"{a}is\"{a}l\"{a} frequency associated with the conductive
temperature profile, ${\bar \nu}$ the molecular viscosity
and ${\bar \kappa}$ the thermal diffusivity. 
These non-dimensional numbers compare the dimensional time scales 
associated with
the dynamics ${\bar t}_{S}= {\bar L}/\overline{\Delta U}$,
the stable stratification ${\bar t}_{N}= 1/{\bar N}$,
the viscous dissipation ${\bar t}_{\nu} = {\bar L}^2/ {\bar \nu}$,
and the thermal diffusion ${\bar t}_{\kappa} = {\bar L}^2/ {\bar \kappa}$.

Due to the horizontal homogeneity,
perturbations are resolved into modes of the form,
\begin{equation} 
w(x,y,z) = \Re (\hat{w}(z) e^{i(k_x x + k_y y - k_x c t)}),
\end{equation}
\noindent
where $\Re ()$ denotes the real part of a complex number,
$\hat{w}(z)$ is the vertical profile of the mode (here 
the vertical velocity), $k_x$ and $k_y$ are the horizontal wavenumbers,
and $c = c_r + i c_i$ is a complex number which real part
$c_r$ is the phase velocity and which imaginary part $c_i$ 
controls the growth or the decay of the mode. If $c_i$ is positive,
$\gamma = k_x c_i$ is the
growth rate of the unstable mode.
Since two-dimensional perturbations contained
in the plane $(x,z)$
are more unstable than three-dimensional ones
(Gage \& Reid 1968),
the spanwise velocity and the wavenumber $k_y$ can be
taken 
equal to zero. Previous investigations on the stability of
parallel shear flow
indicate that the most unstable modes have symmetry properties according
to the symmetry of the basic velocity profile.
Here, due to the antisymmetry of the basic profiles with respect to $z=0$,
we can assume that $\hat{w}(z) = \hat{w}^{*}(-z)$ and
$\hat{\phi}(z) = \hat{\phi}^{*}(-z)$, where $\hat{\phi}^{*}$
is the complex conjugate of $\hat{\phi}$. This implies that $c_r = 0$.
After linearization and some simple manipulations, one gets the two equations,
\begin{equation}\label{eq:wstab}
[\frac{M^2}{Re} - i k_x (U - c) M +i k_x U"] \hat{w} =
Ri k_x^2 \hat{\phi},
\end{equation}
\begin{equation}\label{eq:tstab}
[\frac{M}{Pe} - i k_x (U - c)]\hat{\phi} = \hat{w},
\end{equation}
\noindent
where $\hat{\phi}$ is the vertical profile of temperature disturbance
and $M$ denotes the operator $\frac{d^2}{{dz}^2} - k_x^2$.
The functions $U$ and $U"$ are the basic
velocity profile and its second derivative,
respectively. Together with the 
boundary conditions 
\begin{equation} \label{eq:bc}
\hat{w}, \hat{\phi} \rightarrow 0 \;\; as \;\; z \rightarrow \pm \infty,
\end{equation}
\noindent
Eqs. (\ref{eq:wstab}) and (\ref{eq:tstab}) form an eigenvalue
problem, denoted (E1), which has been solved numerically.

The eigenvalue problem depends on five
parameters $k_x$, $c_i$, $Ri$, $Pe$, $Re$ and may be written
as
${\cal F}(k_x,c_i,Ri,Re,Pe) = 0$.
The hypersurface
of marginal stability separating stable and unstable domains in the parameter
space is given by the condition $c_i = 0$.

\subsection{Numerical model}

The linear system of ordinary differential equations
(\ref{eq:wstab}) and (\ref{eq:tstab}) is solved
with a numerical code that makes use of a
finite-differences technique with deferred correction coupled to a Newton
iteration. The step size can be dynamically refined in the regions where
the eigenfunctions present strong spatial gradients.
 
Boundary conditions are applied at some distance $z_{\rm max}$ from the shear
layer. Then, we increase $z_{\rm max}$ until the eigenvalues 
do no longer depend on the boundary conditions. Depending on the particular
eigenfunction, $z_{\rm max} \approx 20 \; {\rm or} \; 30$ appeared to 
be sufficient to
obtain an accurate solution.
In the inviscid limit, the differential equations are singular at $z=0$
for neutral modes ($c_i = 0$). 
Following Dudis (1974), we overcame this difficulty by solving
the non-singular
problem for $c_i \neq 0$ and then by decreasing 
$c_i$ until the first two digits 
of the eigenvalue
are not affected by the decay of $c_i$. 
These procedures have been successfully tested 
by recovering Dudis results.

\section{Results}

We first
present an overall picture of the
stability properties for
a wide range of Richardson and P\'eclet numbers,
restricting ourselves to
the inviscid case (subsection 3.1).
Then we concentrate on the regime of small P\'eclet number (subsection 3.2).
The growth rates and the spatial structure of the unstable modes are
studied in this regime
(subsection 3.3) as well as the
effect of a large Reynolds number (subsection 3.4).

\subsection{Overall presentation of the stability domains in the inviscid case}

The
marginal stability curves 
are shown in Fig. 1 for a wide range of P\'eclet numbers: 
$Pe = + \infty$, 
$Pe  =1$, 
$Pe  =10^{-1}$, $Pe  =10^{-2}$ and $Pe  =10^{-3}$.
For any fixed value of $Pe$, the right hand side of the
marginal curve corresponds to stable perturbations whereas the left hand
side corresponds to unstable ones. 
The $Pe  = + \infty$ curve 
is the analytical solution $Ri_c = k_x^2 (1-k_x^2)$ 
determined by Drazin (1958) (see also
Chandrasekhar, 1961) in the
non-diffusive limit. Each of the other curves has been drawn 
from about thirty numerically computed triplets ($k_x, Ri, Pe$) 
between $k_x=0.01$ and $k_x=1$.

\begin{figure*}
\resizebox{\hsize}{!}{\includegraphics{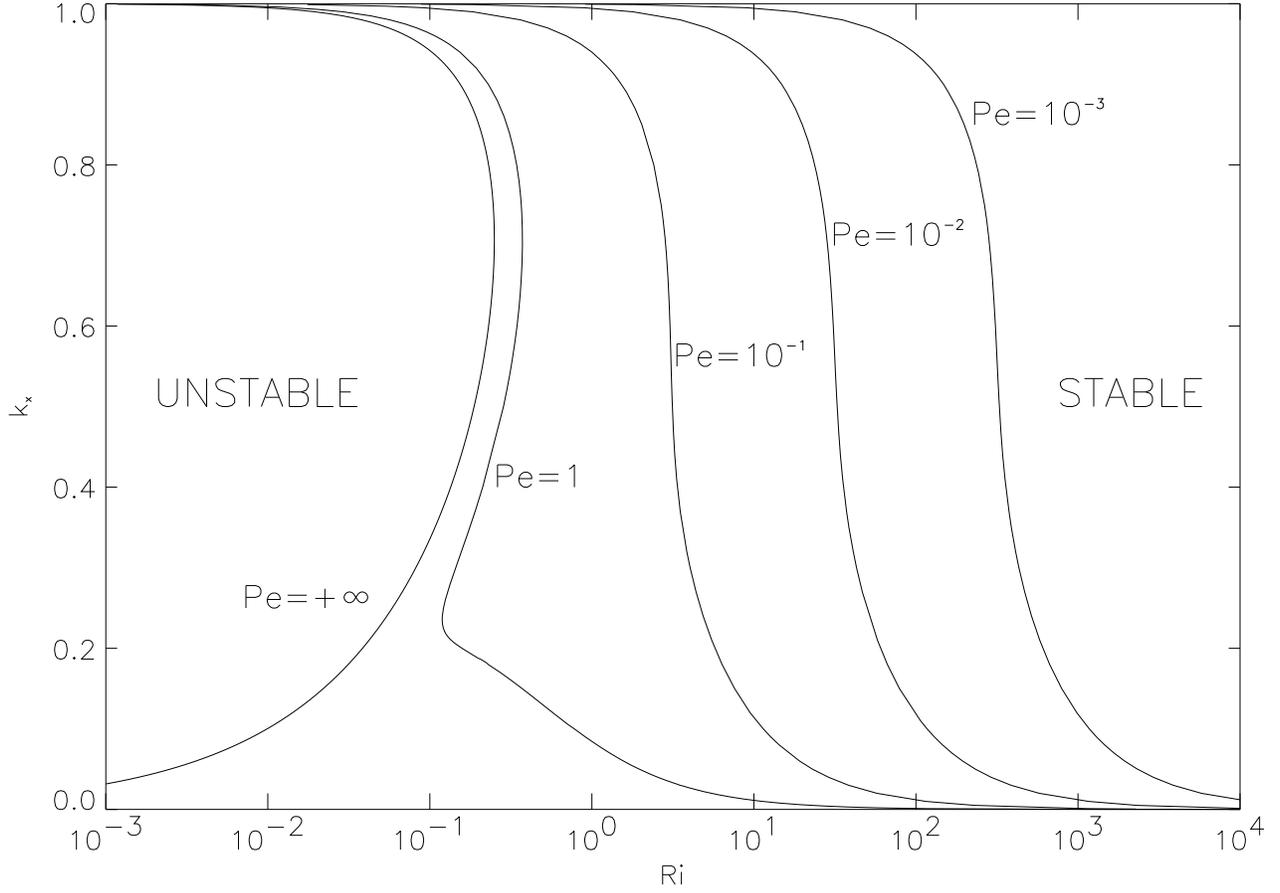}}
\caption{Curves of marginal stability 
separating the stable and unstable domains
in the plane $(k_x , Ri)$, for different values of the P\'eclet number}
\end{figure*}

It can be seen on Fig. 1 that a large thermal 
diffusivity has a destabilizing effect since the unstable domain
increases as the 
P\'eclet number decreases. 
This is easily
explained as follows:
When buoyancy is negligible, the temperature behaves like
a passive scalar and
thermal diffusion has no effect on the dynamics. Then, a first condition
for a significative effect of thermal diffusion is that the flow be affected by the 
stable stratification, that is ${\bar t}_N < {\bar t}_S$ or 
equivalently $Ri >1$. The second condition
is that thermal diffusion reduces the stabilizing effect of the stratification.
This will occur if thermal diffusion acts faster than stable stratification,
that is if ${\bar t}_{\kappa} < {\bar t}_N$. Consequently the
necessary conditions for a significant effect of the thermal diffusion are,
\begin{equation} \label{eq:mio}
{\bar t}_{\kappa} < {\bar t}_N < {\bar t}_S. 
\end{equation}
\noindent
If one considers a shear layer 
characterized by a Richardson number close to unity i.e. such
that ${\bar t}_N \simeq {\bar t}_S$, condition (\ref{eq:mio})
states that thermal diffusion will
reduce the stabilizing effect of stable stratification if 
${\bar t}_{\kappa} < {\bar t}_N \simeq {\bar t}_S$ that is if $Pe <1$. 
Now, if the shear layer is very strongly stratified that is $Ri \gg 1$,
or ${\bar t}_N \ll {\bar t}_S$, condition (\ref{eq:mio})  requires that
${\bar t}_{\kappa} \ll  {\bar t}_S$ or $ Pe  \ll 1$ to destabilize the layer.
These simple arguments account for the mean displacement of the 
neutral curves towards
larger Richardson number, starting from the $Pe = 1$ curve.
Thus, condition (\ref{eq:mio}) can be said to 
explain the global behaviour of the stability 
picture shown in Fig. 1.

However, another point apparent on Fig. 1 is that
large horizontal wavelengths are stable for $Pe = + \infty$ but become
unstable as $Pe$ decreases. They become even the most unstable
modes at fixed values of $Pe$! 
This property is harder to understand
because thermal diffusion acts primarily
on small scale perturbations.

We give here a partial answer by showing how perturbations with
large horizontal wavelength can be strongly affected by 
thermal diffusion:
Again, in order to weaken the stable stratification, thermal exchanges
must act before stable stratification. This condition has been 
previously expressed by ${\bar t}_{\kappa} < {\bar t}_N$ but this inequality
does not 
take into account the anisotropy of the buoyancy force.
Actually, the effect of stable stratification tends to vanish
when motions become more and more horizontal. By contrast, 
thermal diffusion remains efficient for these motions since
predominantly
horizontal motions must vary on a relatively small vertical scale to ensure
mass conservation ($\nabla \cdot {\bf u} = 0$).
Therefore, thermal diffusion tends to be more efficient than 
stable stratification 
for predominantly horizontal motions.

This reasoning can be formalized by considering infinitesimal
perturbations of the form $w \propto \exp [i(k_x x + k_z z)]$ in a 
linearly stratified atmosphere at rest. The stable stratification time scale 
then corresponds to the inverse of 
gravity waves frequency,
\begin{equation} \label{eq:bou}
 t_{N}(k_x, k_z) =
\sqrt{k_x^2 + k_z^2}/ k_x \sqrt{Ri}, 
\end{equation}
\noindent
whereas the thermal diffusion time scale is
simply,
\begin{equation}
t_{\kappa}(k_x, k_z) =  Pe/(k_x^2 + k_z^2).
\end{equation}
\noindent
Comparing both time scales shows that, whatever the value 
of the P\'eclet number, 
thermal diffusion acts much faster than stable stratification
when $\mid k_x/k_z\mid = \mid w/u \mid$ goes to zero, that is
when motions are predominantly horizontal. 

To show that the motions associated
with the marginal modes at large horizontal wavelengths 
are in this case, we
calculated the square root of 
the ratio between the vertical and horizontal kinetic energy
of the modes. The ratio is given 
by $E_{w}/E_{u}$ where,
\begin{equation} \label{eq:rat}
E_w = \frac{\int _{-\infty}^{+\infty}  \int_{0}^{2\pi/k_x}   w^2 \;\; dx dz}
{\int _{-\infty}^{+\infty}  \int_{0}^{2\pi/k_x}   ({w}^2 + u^{2}) \;\; dx dz},
\end{equation}
\noindent
and $E_u$ is defined accordingly. It has been
determined
for marginal modes 
taken along the $Pe =1$ neutral 
curve and the results are presented on Fig. 2. It clearly 
appears that disturbances 
with large horizontal wavelengths are associated with strongly 
anisotropic, predominantly horizontal 
motions. Actually, in the limit of small $k_x$,
the eigenfunctions $\hat{w}$ and $\hat{\phi}$
do not vary anymore so that the vertical scale height of the disturbance
remains fixed
while its horizontal wavelength increases. 
This explains why thermal diffusion can affect 
their stability, regardless of the value of the P\'eclet number.
\begin{figure}
\resizebox{\hsize}{!}{\includegraphics{9010.f2}}
\caption{Square root of the ratio between the vertical and horizontal kinetic
energy of neutral modes taken 
along the $Pe = 1$ marginal stability curve. This figure
shows that the motions associated with these modes become more and more 
horizontal as $k_x$ goes to zero}
\end{figure}
What is still unclear is why they are the most unstable 
disturbances. 
We shall come back to this point
at the end of the next subsection.

\subsection{The small P\'eclet number regime}

In the following, we concentrate our analysis on the small P\'eclet number
regime which is most relevant for stellar astrophysics.
As suggested by Dudis' work, we plot once again 
the neutral curves corresponding to
$Pe= 1, 10^{-1}, 10^{-2}, 10^{-3}$ but this time as a function of $Ri Pe$.
The outcome is displayed on Fig. 3. We observe that three of the
marginal stability curves
nearly collapse into a single curve. While differences between the 
$Pe= 10^{-2}$ and $Pe= 10^{-3}$ curves 
are not distinguishable,
the $Pe =10^{-1}$ curve slightly deviates from those.

\begin{figure*}
\resizebox{\hsize}{!}{\includegraphics{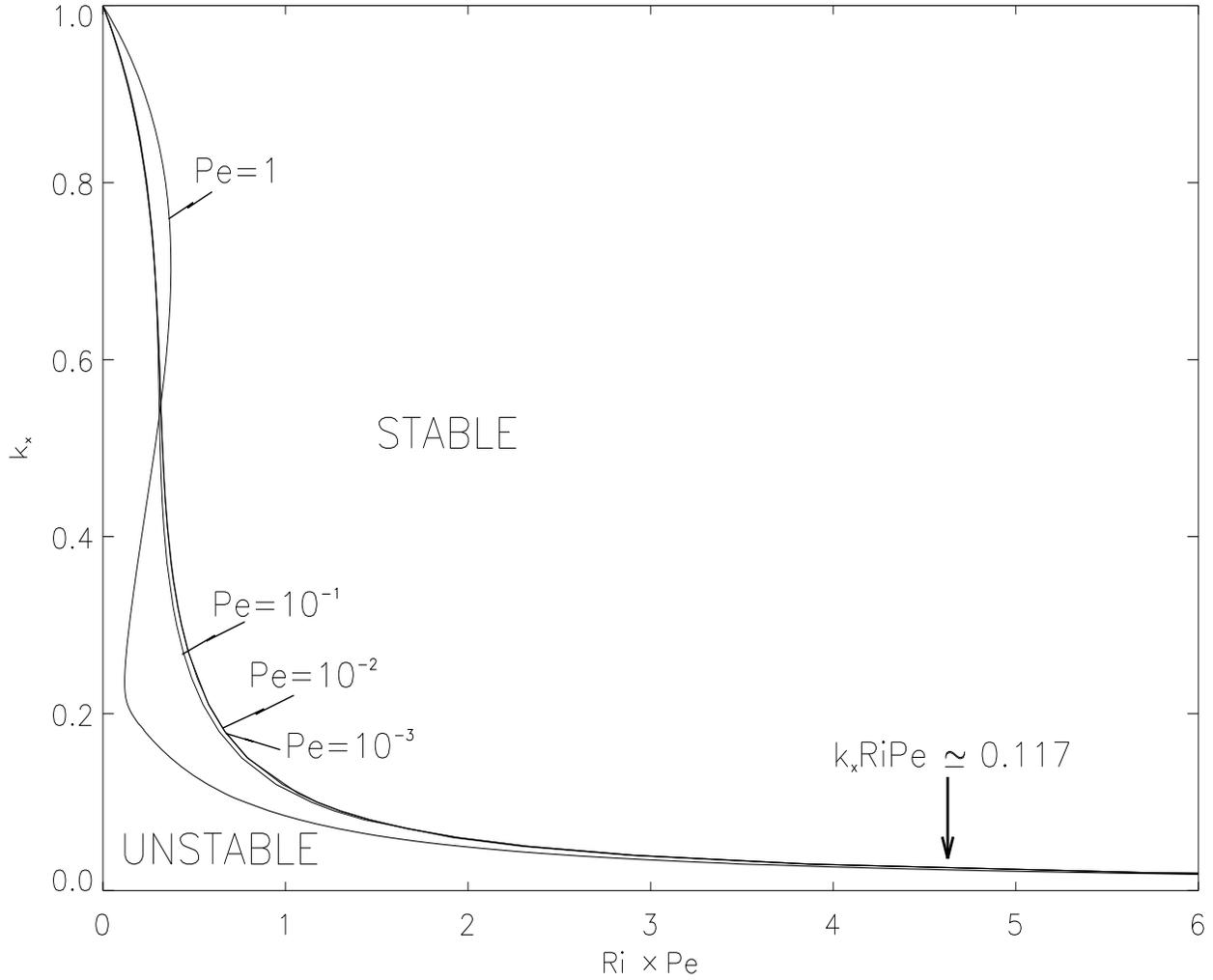}}
\caption{The marginal stability 
curves shown on Fig. 1 are now plotted as a function
of $Ri Pe$. The collapse of the $Pe = 10^{-2}$ and $Pe = 10^{-3}$ curves
suggests the existence
of an asymptotic regime in the limit of small P\'eclet number.
Moreover, all the curves tends towards
the curve $k_x Ri Pe \simeq 0.117$ as $k_x$ goes to zero,
suggesting the existence of another asymptotic regime
in the limit of small horizontal wavenumber}
\end{figure*}

This remarkable property strongly suggests the existence of an asymptotic state for 
small P\'eclet numbers in which the governing equations only depend on the
non-dimensional number, $Ri Pe$. Dudis (1974) had already found
that the eigenvalue 
problem for neutral modes depends only on the product $Ri Pe$
if the mean flow advection is neglected in Eq.
(\ref{eq:tstab}).
Using asymptotic expansions and starting from the Boussinesq equations, 
Ligni\`eres (1999) recently 
obtained a general form of the asymptotic equations
in the limit of small P\'eclet numbers.
It is shown that, if ${\bf u}$ and $\theta$ behave like Taylor series
for small P\'eclet numbers (${\bf u} = {\bf u_0} + Pe {\bf u_1} + ...,
\theta = \theta_0 + Pe \theta_1 + ...$), 
the solutions of the Boussinesq equations 
are identical to the solutions of the following system up to the
first order in $Pe$:
\begin{equation} \label{eq:vela0}
\frac{\partial {\bf u}}{\partial t} + {\bf u}\cdot \nabla {\bf u} =
- \nabla p + R \psi {\bf e}_z + \frac 1{Re} \nabla ^2{\bf u},
\end{equation}

\begin{equation} \label{eq:tempa0}
\nabla ^2 \psi = w,
\end{equation}

\begin{equation}  \label{eq:diva0}
\nabla \cdot  {\bf u}=0,
\end{equation}
\noindent
where $\psi = \theta / Pe $ are the temperature deviations scaled by $Pe$
and
\begin{equation}
R = Ri Pe = \frac{{\bar t}_S {\bar t}_{\kappa}}{{\bar t}_N^2}.
\end{equation}
\noindent
In this context, 
the linear equations governing the shear layer stability become:
\begin{equation}\label{eq:wstabpe}
[\frac{M^2}{Re} - i k_x (U - c) M +i k_x U"] \hat{w} =
R k_x^2 \hat{\psi},
\end{equation}
\begin{equation}\label{eq:tstabpe}
M \hat{\psi} = \hat{w},
\end{equation}
\noindent
which, as expected, only depend on the non-dimensional 
numbers, $R = Ri Pe$ and $Re$.
The eigenvalue problem formed by the above equations and the 
boundary conditions (\ref{eq:bc})
has been solved numerically 
for neutral ($c =0$) and inviscid ($Re = + \infty$) modes.
We found that the resulting 
marginal stability curve can not be distinguished from
the $Pe =10^{-2}$ and $Pe =10^{-3}$ curves 
obtained in the context of the Boussinesq equations.

The existence of this asymptotic regime
has various interests for the stability analysis.
First, we can concentrate on the asymptotic limit and do not have 
to repeat the analysis for different
values of the P\'eclet number. Second, it turns out that
the stability is no longer
governed by three processes (shear, stable stratification and thermal diffusion)
but only by two, the shear destabilization and a stablizing process 
combining the stable stratification and thermal diffusion effects.

To understand why, let us consider
the differences
between the 
heat conservation of the 
Boussinesq equations (Eq. (\ref{eq:temp1})) and the simplified 
heat balance of the small-P\'eclet-number approximation (Eq. (\ref{eq:tempa0})).
This simplified balance occurs because, unlike other advection terms, 
the vertical advection against the mean stratification
does not vanish when the high thermal diffusion strongly reduces
the temperature deviations.
Indeed, as fluid parcels go up (or down) in a mean temperature gradient,
the amplitude of the associated temperature deviations tends to increase
continuously. While this process acts as a source of temperature deviations, 
thermal diffusion tends to smooth them out.
Eq. (\ref{eq:tempa0}) represents
the balance between both processes.
In this state, vertical advection against mean temperature gradient
and thermal diffusion no longer act individually
to determine the amplitude of the
buoyancy force; the process which combines their action
will be called 
the small-P\'eclet-number buoyancy in
the following of the paper.
Its elementary properties have been analyzed in Ligni\`eres (1999) and
are briefly summarized here.
It is purely dissipative in the sense that
the work of the buoyancy force integrated over the whole domain is always
negative, provided temperature deviations vanish on the outer boundaries.
Its time scale
is ${\bar t}_{B} = {\bar t}^{2}_{N} /{\bar t}_{\kappa}$,
although the process being strongly anisotropic, the time scale
of the small-P\'eclet-number buoyancy is
better described by, 
\begin{equation} \label{eq:tpe}
t_{B}(k_x, k_z) = \frac{\left(k_x^2 + k_z^2 \right)^{2}}{R k_x^2},
\end{equation}
\noindent 
which corresponds to
the time scale necessary to dissipate an infinitesimal perturbation of the form 
$w \propto \exp [i(k_x x + k_z z)]$ in the atmosphere at rest.

Coming back to the stability analysis, we shall now compare the stabilizing
effect of the small-P\'eclet-number buoyancy with the destabilizing
one of the shear.

Let us consider 
the instability of large horizontal wavelength disturbances.
In the previous subsection, we have shown that
the vertical length scale of these disturbances
remains fixed as their horizontal wavenumber goes to zero.
Then, according to expression (\ref{eq:tpe}), 
the small-P\'eclet-number buoyancy time scale increases like $1/k_x^2$
in the same limit.
By comparison, the time scale of the instability (the inverse of the 
growth rate in a fluid of constant density) is known to be proportional 
to $1/k_x$ for small $k_x$
(Drazin 1981).
Consequently, the stabilizing effect decreases faster than the destabilizing one
as $k_x$ vanishes. This explains why predominantly horizontal perturbations
are unstable and why they are increasingly unstable as $k_x$ tends to zero.

It is worth mentioning 
the difference with the non diffusive case ($Pe = + \infty$). 
In this case also, the stable stratification is inefficient for
predominantly horizontal disturbances. But, by contrast with
the diffusive situation, the stable stratification 
time scale increases as $1/k_x$ (see Eq. (\ref{eq:bou})),
that is as fast as the instability time scale.
The result is that disturbances with
large horizontal wavelength are stable (see the $Pe = + \infty$ neutral-curve
on Fig. 1). 
We therefore conclude that the shapes of the non-diffusive and highly diffusive 
marginal stability curves differ because the non-diffusive buoyancy
has a stronger effect on predominantly horizontal motions than
the small-P\'eclet-number buoyancy. In other words, the action of
the 
small-P\'eclet-number buoyancy is more anisotropic than that of the
non-diffusive buoyancy,
and this triggers the instability of predominantly horizontal motions.

The small-P\'eclet-number buoyancy allows also to understand why,
as indicated in Fig. 3, the marginal stability curves converge
towards $k_x R = 0.117$ when $k_x$ goes to zero.
Since stabilizing and destabilizing effects
balance each other for marginal modes, the marginal curve
can be characterized by an equality between 
the dynamical time scale
and the small-P\'eclet-number buoyancy time scale. 
In the limit of vanishing $k_x$, this reads:
\begin{equation}
\frac{1}{k_x} = \frac{k_z^4}{k_x^2 R},
\end{equation}
\noindent that is,
\begin{equation}
k_x R = constant,
\end{equation}
\noindent since the vertical scale height of these modes remains fixed.

Although outside the regime of small P\'eclet numbers,
we observe that
the $Pe=1$ marginal curve also converges towards $k_x R = 0.117$
for small $k_x$. This is due to the fact that
modes associated with predominantly
horizontal motions 
satisfy the simplified heat equation
(\ref{eq:tempa0}) whatever the value of the P\'eclet number.
In the limit of small $k_x$,
the advective term $U \partial \theta/ \partial x$ vanishes
naturally
while
the temporal variation $\partial \theta/\partial t$
is also negligible because,
as we have seen before,
thermal diffusion dominates stable stratification for any value of the
P\'eclet number provided the motions are sufficiently horizontal.

From a mathematical point of view, the convergence towards $k_x R = 0.117$
means
that the eigenvalue problem has reached an asymptotic regime
in the limit of small $k_x$.
Effectively,
when terms proportional to
$k_x^2$ are neglected
in Eqs. (\ref{eq:wstabpe}) and (\ref{eq:tstabpe}), the asymptotic
eigenvalue
problem depends only on the parameter $k_x R$.
Jones (1977) had already obtained this asymptotic eigenvalue problem, 
starting from Eqs. (\ref{eq:wstab}) and (\ref{eq:tstab})
and considering the double limit $k_x \rightarrow 0$
and $Ri \rightarrow +\infty$. He found a slightly smaller value, 
$k_x R =0.086$,
most probably because of the difference in the boundary conditions. In his
study, the
requirement that perturbations vanish at $z=\pm 4$
decreases a little bit the domain of
instability.

\subsection{Growth rates in the small P\'eclet number regime}

In the unstratified case, it is well known that the growth rate of the
instability
is maximum for a particular mode 
($\gamma_0 \simeq 0.189$  at $k_x \simeq 0.447$, see for example Drazin 1981).
In the same way,
one would like to know whether, for a fixed value of $R = Ri Pe$, a particular 
mode will grow faster than the others. In that case, a relevant issue
will be to quantify the reduction of the 
maximum growth rate due to the
small-P\'eclet-number buoyancy.

Contours of constant growth rate have been determined inside the unstable domain
and are presented on Fig. 4. Unsurprisingly, 
the growth rates decrease when $Ri Pe$ increases, 
that is when the stabilizing effect of the buoyancy force becomes stronger.
For example at $ Ri Pe \simeq 5$, the maximum growth rate 
is $\simeq 0.001$, that is smaller by more than two order of magnitude than
the maximum growth rate of the unstratified case.

\begin{figure}
\resizebox{\hsize}{!}{\includegraphics{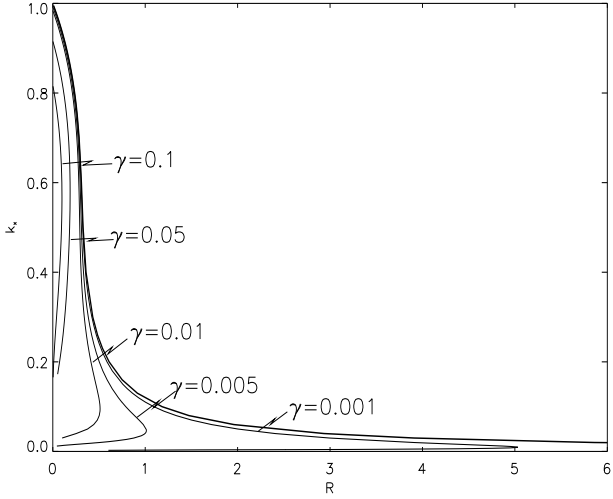}}
\caption{Growth rate isocontours in the limit of small P\'eclet numbers. The
bold curve corresponds to marginal stability ($\gamma =0$). Note that the
maximum growth rate in the unstratified case is $\gamma_0 \simeq 0.189$}
\end{figure}

It can be inferred by looking at Fig. 4 that
a maximum growth rate exists 
for each fixed value of $R = Ri Pe$. 
This maximum value denoted ${\gamma}_{\rm max}$ 
is shown on Fig. 5 as a function
of $R$. It appears that
a rapid decrease of
the maximum 
growth rate is followed by a much slower one after an abrupt change of slope
at $R \simeq 0.253$.
This change of behaviour
is due to the existence of two different types of
modes as supported by
Fig. 6 where, $k_x^{\rm max}$,
the horizontal wave number of the most
amplified disturbance is plotted against ${\gamma}_{\rm max}$.

Figs. 5 and 6 show that a first type of mode characterized by 
horizontal wavenumbers close to $0.5$
is dominant in the regime $R < 0.253$
whereas a second type of mode with much smaller wavenumbers becomes dominant once $R> 0.253$. 
The "jump" between both types of modes occurs at $R \simeq 0.253$ when
the curve $\gamma =  f (k_x)$ has two equal local maxima.

\begin{figure}
\resizebox{\hsize}{!}{\includegraphics{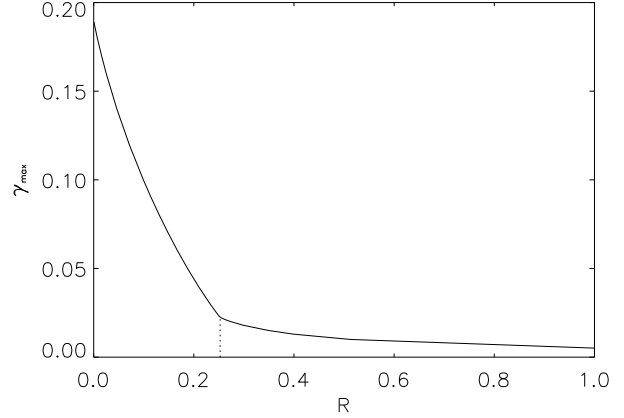}}
\caption{Maximum growth rate as a function of $R = Ri Pe$.
Only the range $0<R<1$ is represented here}
\end{figure}

\begin{figure}
\resizebox{\hsize}{!}{\includegraphics{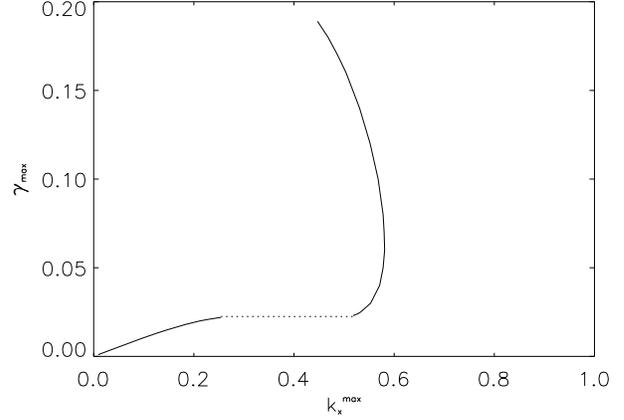}}
\caption{Horizontal wavenumber of the most unstable modes}
\end{figure}

The vertical structure of both types of modes has been analyzed in order
to characterize their geometrical properties.
For the modes which dominate at small values of $R$,
the vertical decay outside the shear layer is quite rapid since 
their vertical scale height
varies between $\bar L$ and $2 \bar L$. 
These modes can be said to be isotropic as their vertical and horizontal length scale are of the same order.

By contrast, the modes which dominate for large values of $R$ are strongly
anisotropic and become more and more so as $R$ increases.
Effectively, we found that their vertical scale height remains fixed to $\simeq 5.56 {\bar
L}$
while their horizontal wavelength ranges from $24 \bar L$ to $+ \infty$
as $R$ increases.
Another property of this type of modes is that their growth rate and
horizontal wavenumbers are related to $R$ through scaling laws.
This can be seen on Fig. 7 where the relation between $\gamma_{\rm max}$ and
$R$ is plotted again using log-log coordinates 
and including large
values of $R$.  
\begin{figure}
\resizebox{\hsize}{!}{\includegraphics{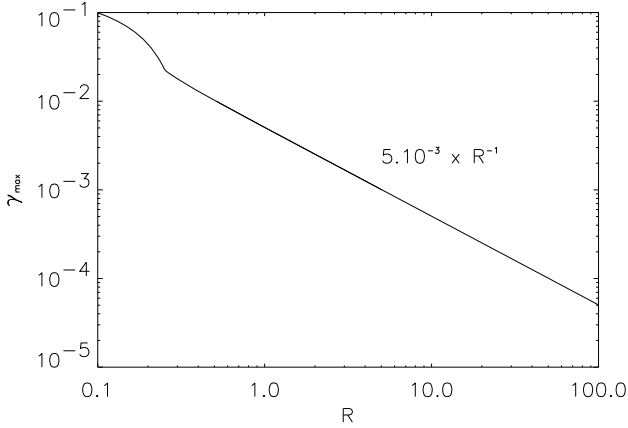}}
\caption{Maximum growth rate as a function of $R = Ri Pe$ in log-log
coordinates
}
\end{figure}
The scaling law,
\begin{equation} \label{eq:sca1}
\gamma_{\rm max} \simeq 5.\times 10^{-3}  R^{-1},
\end{equation}
\noindent
is rapidly established as $R$ increases above $0.253$.
In the same way, it is found that 
the wavenumber of the most amplified mode
and $R$ are related by,
\begin{equation} \label{eq:sca2}
k_x^{\rm max} \simeq 4.86 \times 10^{-2}  R^{-1}.
\end{equation}
\noindent
These scaling laws are due to the fact that the eigenvalue problem
reaches an asymptotic regime for small horizontal wavenumbers.
Indeed,
in the limit $k_x \rightarrow 0$, the
eigenvalue problem for
non-neutral modes ($k_x c_i = \gamma \neq 0$)
is governed by two parameters only,
$a = k_x R$ and $b = \gamma / k_x$.
Relations (\ref{eq:sca1}) and (\ref{eq:sca2}) then show that
the most unstable mode 
corresponds to a particular solution of the asymptotic
eigenvalue problem.

\subsection {Effect of a large Reynolds number in the small 
P\'eclet number regime}

For a constant density fluid,
the Kelvin-Helmholtz instability is said to be inviscid because
viscosity does not play any role in the mechanism which triggers
the instability. Viscous dissipation
can still
influence the
stability but, at least for
the hyperbolic-tangent shear layer, its effect is
rapidly negligible as the Reynolds number increases.
This is not the case in a highly diffusive atmosphere since we found that
viscous dissipation has a significant 
influence even if
the Reynolds number is very large. This point is illustrated by Fig. 8, where
the inviscid neutral curve is compared with the neutral curve calculated 
at $Re=1000$.
\begin{figure}
\resizebox{\hsize}{!}{\includegraphics{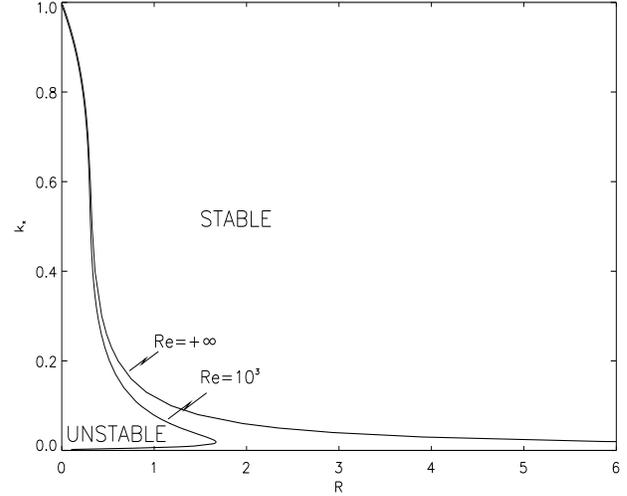}}
\caption{Curves of marginal stability for $Re=1000$ and $Re= + \infty$
in the small-P\'eclet-number regime}
\end{figure}
The figure shows that viscosity essentially comes into play 
by stabilizing the large scale 
predominantly horizontal disturbances, its 
influence on the other type of perturbations
being much weaker.

This can be explained as follows: 
for wavelengths of the order of the shear layer
thickness, the viscous time scale is much larger
than the other time scales. However, as $k_x$ goes to zero, both the time
scale of the small-P\'eclet-number buoyancy and the time scale of 
the instability become infinite while the 
viscous time scale tends towards a constant value,
\begin{equation}
t_{\nu} = H_z^2 R_{e},
\end{equation}
\noindent where $H_z$ denotes the vertical length scale of the
predominantly horizontal modes. Then, at some point, viscous dissipation 
becomes dominant and stabilizes
the perturbations.

The 
critical values of $R$ and $k_x$ have been determined
for two other Reynolds numbers, $Re=1100$
and $Re=1500$. They staisfy
the following relations,
\begin{equation} \label{eq:re1}
\frac{R_{\rm crit}}{Re} = 1.67 \times 10^{-3} \;\;\; Re k_x^{\rm crit} = 19,
\end{equation}
\noindent so that the stability criterion is,
\begin{equation} \label{eq:re2}
Ri Pr = \frac{R}{Re} > 1.67 \times 10^{-3},
\end{equation}
\noindent for the three Reynolds numbers considered.

There are two different ways of recovering the
relations (\ref{eq:re1}) by simple arguments.
First, by saying that all
perturbations are stable when viscous damping balances 
the growth of the most unstable mode
in the inviscid case. Equalizing 
the viscous time 
scale $t_\nu$ with 
the inverse of the maximum
growth
rate $1/\gamma_{\rm max}$ yields 
$R/Re = constant$. The second scaling law
(\ref{eq:sca1}) shows that the
corresponding wavenumber satisfies $k_x Re = constant$.

The same relations can be obtained without knowing the expression of the
maximum growth rate. We first define a stabilizing time
scale which includes the effects of the small-P\'eclet-number
buoyancy and the viscous dissipation as,
\begin{equation}
t_{\rm stab} = \frac{\left(k_x^2 + k_z^2 \right)^{2}}{R k_x^2 +
\frac{\left(k_x^2 + k_z^2 \right)^{3}}{Re}}.
\end{equation}
\noindent
It corresponds to the damping time of an
infinitesimal perturbation proportional
to $\exp [i(k_x x + k_z z)]$ in a quiescent atmosphere
described by the small-P\'eclet-number approximation.
Then, the marginal stability curve can be
characterized by an equality
between this time scale
and
the instability time scale.
In the limit of small $k_x$, it turns out that
the
function $R(k_x)$ possesses a maximum $R_{\rm max}= Re k_z^2 /4$,
reached at $k_x Re =k_z^2$. The above relations follow 
since the vertical scale height of these modes remains fixed.

For smaller Reynolds numbers ($25 \leq Re \leq 150$), Jones (1977) found
a criterion similar to (\ref{eq:re2}), although the constant is somewhat
higher 
$Ri Pr > 7 \times 10^{-3}$. This difference can
be attributed to the
boundary conditions. 

\section{Summary and discussion}

The linear stability of a hyperbolic-tangent shear layer in a stably stratified
atmosphere has been investigated for a wide range of Richardson,
P\'eclet and Reynolds numbers. 
Emphasis has been put on the regime of small P\'eclet numbers (i.e. high thermal
diffusivity) relevant for stellar interior dynamics. Note that this regime is 
also important for the upper atmosphere of the Earth and Venus
as Townsend (1958) and Dudis (1974) pointed out.

In the inviscid case, we found that the shear layer is always unstable
provided the horizontal wavelength of the perturbations is not bounded.
It should be specified
that the growth rate
goes to zero as the horizontal wavelength goes to infinity.
Viscous dissipation stabilizes this type of perturbation
and so introduces
a critical value of $R = RiPe$ above which the shear layer is stable.

The analysis of the growth rates in the inviscid small P\'eclet number regime
revealed the existence of two different types of modes. A first type,
isotropic and relatively fast growing, is dominant in the range 
$0 < R=RiPe < 0.253$. By relatively fast, we mean that the growth rate is 
always larger than $0.1 \gamma_0$, where $\gamma_0$ is the
maximum growth rate in the unstratified case.
For larger values of $R$,
the most unstable modes are predominantly horizontal. Their
growth
rates continue to decrease, following the
scaling law 
${\bar {\gamma}}_{\rm max} = 5. \times 10^{-3}
{Ri}^{-1} {\bar t}_{\kappa}^{-1}$,
in dimensional units.

Perhaps, the most striking result is the existence of an asymptotic regime
where the mathematical and physical analysis of the stability
is considerably simplified. The physical simplification comes
from the fact that the buoyancy force is no longer determined by two
independent processes, temperature advection
in the stratified atmosphere
and thermal diffusion, but rather by a unique process combining both.
The effect of this so-called small-P\'eclet-number
buoyancy
turns out to be purely dissipative and strongly anisotropic.
Comparing its stabilizing effect to the destabilizing
Kevin-Helmholtz mechanism
leads to a straightforward interpretation of the
stability properties.

It should be noted that
the small-P\'eclet-number buoyancy characterizes
the response of a highly diffusive 
stably stratified atmosphere to any sufficiently slow motions. 
As such, it can potentially be applied
to other types of motions than those driven by a parallel shear flow.
This includes for example other types of double diffusive instabilities
like the Goldreich-Schubert instability or the thermohaline instability.
We note in particular that the most unstable modes of the latter correspond to
long thin columns. This is fully compatible with
the properties of the small-P\'eclet-number buoyancy since the 
associated stabilizing time scale (\ref{eq:tpe}) becomes infinite for this
type of motions 
($k_x \rightarrow +\infty$ and $k_z$ remaining fixed).
Work is in progress to assess the usefulness of the small-P\'eclet-approximation
in this context.

Strictly speaking, the results presented here 
are only valid for the particular flow considered and
assuming that the perturbations are infinitesimal. We expect nevertheless that similar results
would be obtained for any shear layer subject to the Kelvin-Helmholtz instability, that is
when the vorticity profile possesses an extremum (Drazin, 1981). But, in any case,
our results can not be directly extended to the other
velocity profiles and more importantly if the basic flow is perturbed by finite amplitude
disturbances. Effectively, since Reynolds'
experiments on
flows in a pipe (Reynolds 1883), it is well known that finite 
amplitude perturbations are able to trigger
shear layer instabilities when infinitesimal ones can not. 
In this context the criterion we obtained 
appears as a sufficient - but not necessary -
condition for instability. 

In principle, it can be applied as such, to shear layers inside stellar
radiative zones. However, taking typical solar 
values for the thermal diffusivity and the Brunt-V\"{a}is\"{a}l\"{a}
frequency ($\kappa = 10^7  \mbox{\rm cm}^{2}  \mbox{\rm s}^{-1}$ 
and $N= 10^{-3} \mbox{\rm s}^{-1}$), 
it is found
that shear layers unstable according to this criterion would have vertical
extent much smaller than the vertical resolution of helioseismology data ($L/R_{\sun} < 10^{-5}$). To give
an example, if one assumes that the shear across the solar tachocline can be modeled by a
hyperbolic-tangent profile entirely embedded in the radiative zone,
the associated Richardson number
would be about $2.5 \times 10^3$ and the P\'eclet number about $10^6$
(we took $5$ percent of the solar radius for the tachocline thickness
and $\Delta \Omega = 10^{-6} \mbox{\rm s}^{-1}$ for the differential rotation
across the tachocline).
Then, our criterion 
would predict stability
because the smallest unstable 
horizontal wavelength according to $k_x = 0.117/ Ri Pe$
is much larger than the solar radius. Note again 
that shears across the tachocline could
be subjected
to finite amplitude instabilities
as discussed in Michaud \& Zahn (1998).

For the time being, 
one relies on the criterion established by Zahn (1974) to decide about 
the non-linear stability of a shear layer.
This criterion is not in contradiction
with our results. But this consistency check is not sufficient to prove its validity.
Interestingly enough, the 
energetic consideration made by Maeder (1995) have shown 
that the criterion is consistent energetically. However, 
in the derivation by Zahn (1974) and 
by Maeder (1995), perturbations 
with a velocity scale $u$ and a length scale $l$ 
are introduced ab initio and it is not known whether such perturbations 
correspond to possible 
solutions of the equations of motion. 
Therefore, this criterion can be interpreted as a necessary 
condition for instability. 

Further investigations are clearly needed to better constrain this criterion. 
In a geophysical context, numerical 
simulations of stably stratified sheared turbulence
have been able to find
critical Richardson numbers
for which turbulence neither grows nor decays
(see a review by Schumann, 1996). 
Such simulations should be 
performed at smaller P\'eclet numbers to approach stellar conditions
and, as pointed out by Ligni\`eres (1999), the small-P\'eclet-number
approximation would be very useful in this context.
Another possible approach is
to assume that finite amplitude disturbances provoke a defect of the 
basic profile and then to analyze the 
linear stability of the perturbed velocity profile as a function 
of the defect amplitude. This method has been used by
Dubrulle \& Zahn (1991) for 
the unbounded Couette flow in a constant density
fluid and could be extended
to the stably stratified case at small 
P\'eclet number.

As quoted in the introduction, beyond the stability of shear layer, the final objective
is to estimate the angular momentum and chemical element transport driven by 
shear layer turbulence. Assuming that small scale turbulence behaves like a diffusion process, 
Zahn (1992) proposed an expression for the turbulent viscosity
on the basis of arguments used to derive 
the stability criterion. This expression reads,
\begin{equation} \label{eq:tvis}
{\nu}_{t} = constant \times {\bar \kappa} 
{\left( \frac{{\bar r}}{{\bar N}} \frac{d{\bar \Omega}}{d{\bar r}} \right)}^2
\propto {\bar \kappa} Ri_g^{-1},
\end{equation}
\noindent where $Ri_g$ is the Richardson number formed with the local velocity
gradient ${\bar r}
d{\bar \Omega} /d{\bar r}$ and the local Brunt-V\"{a}is\"{a}l\"{a} frequency.
Clearly our linear stability analysis gives no clues about the 
non-linear processes governing turbulent transport. Nevertheless, as
the instability
mechanism controls the energy injection into turbulence,
it is natural to suppose that changes in the stability properties 
have a direct impact on turbulent transport. 
On dimensional grounds, we can form a vertical transport coefficient using the 
growth rate and the vertical length scale characterizing the most unstable 
mode is 
${\bar {\gamma}_{\rm max}} {\bar H_z}^{2}$.
For large values of $R=RiPe$ ($R > 0.253$), ${\gamma}_{\rm max}$
follows the scaling law (\ref{eq:sca1}) whereas
the typical vertical length of the most unstable mode remains fixed, so that
${\bar {\gamma}_{\rm max}} {\bar H_z}^{2} \propto {\bar \kappa} Ri^{-1}$.
This exactly 
corresponds to the scaling of the turbulent viscosity (\ref{eq:tvis}).
Such a close agreement is not found in the regime of small $RiPe$. 
However, the variation of the maximum growth with $RiPe$ and
the expression of the turbulent viscosity are still compatible.
Actually, the main difference comes from the very existence of two different
regimes 
which are not contained in the expression of the turbulent viscosity
(\ref{eq:tvis}).
One corresponds to rapid 
isotropic motions on the length scale of the shear layer,
the other to slow nearly horizontal motions. 

\begin{acknowledgements}
We are very grateful to Yin-Ching Lo for his careful reading
of the manuscript.
\end{acknowledgements}


\begin{thebibliography}{}
\bibitem [1961]{chan} Chandrasekhar S., 1961, Hydrodynamic and hydromagnetic
stability. Clarendon Press.
\bibitem [1958]{dra0} Drazin P.G., 1958, J. Fluid Mech. 4, 214
\bibitem [1981]{dra1} Drazin P.G., Reid, W.H., 1981, 
Hydrodynamic stability. Cambridge University Press.
\bibitem [1991]{dub} Dubrulle B., Zahn J.-P., 1991, J. Fluid Mech. 231, 561 
\bibitem [1974]{dudis} Dudis J.J., 1974, J. Fluid Mech. 64, 65
\bibitem [1968]{gag} Gage K.S., Reid W.H., 1968, J. Fluid Mech. 33, 21 
\bibitem [1977]{jon} Jones C.A., 1977, Geophys. Astrophys. Fluid Dyn. 8, 165
\bibitem [1999]{lig} Ligni\`eres F., 1999, A\&A in press
\bibitem [1995]{mae} Maeder A., 1995, A\&A 299, 84
\bibitem [1961]{mil1} Miles J.W., 1961, J. Fluid Mech. 10,496 
\bibitem [1986]{mil2} Miles J.W., 1986, J. Fluid Mech.
\bibitem [1998]{mich} Michaud G., Zahn J.-P., 1998, Theoret. Comput. Fluid
Dynamics 11, 183
\bibitem [1883]{rey} Reynolds O., 1883, Phil. Trans. Roy. Soc. 174, 935
\bibitem [1996]{schu} Schumann U., 1996, Dynamics Atmos. Oceans 23, 81
\bibitem [1960]{spi} Spiegel E.A., Veronis G., 1960, ApJ 131, 442
\bibitem [1958]{tow} Townsend A.A., 1958, J. Fluid Mech. 4, 361
\bibitem [1974]{zan} Zahn J.-P., 1974, in:
Ledoux et al. (eds.), Stellar instability and evolution, p. 185
\bibitem [1992]{zan2} Zahn J.-P., 1992, A\&A 265, 115
\end{thebibliography}
\end{document}